\documentclass[aps,prc,10pt,showpacs,showkeys,twocolumn,groupedaddress]{revtex4-1}
\usepackage{epsfig}
\usepackage{graphicx}
\usepackage{graphics}
\usepackage{xspace}
\usepackage[dvips]{color}
\usepackage{latexsym}
\usepackage{mathrsfs}
\usepackage{url}
\newcommand{\inieq}{\begin{eqnarray}}            
\newcommand{\fineq}{\end{eqnarray}}            
\newcommand{\be}{\begin{equation}}
\newcommand{\ee}{\end{equation}}
\newcommand{\ba}{\begin{eqnarray}}
\newcommand{\ea}{\end{eqnarray}}

\begin{document}
\title{Relativistic descriptions of final-state interactions
in neutral-current neutrino-nucleus scattering at MiniBooNE
kinematics.}
\author{Andrea Meucci} 
\affiliation{Dipartimento di Fisica Nucleare e Teorica, 
Universit\`{a} degli Studi di Pavia and \\
INFN,
Sezione di Pavia, via A. Bassi 6, I-27100 Pavia, Italy}
\author{Carlotta Giusti}
\affiliation{Dipartimento di Fisica Nucleare e Teorica, 
Universit\`{a} degli Studi di Pavia and \\
INFN,
Sezione di Pavia, via A. Bassi 6, I-27100 Pavia, Italy}
\author{Franco Davide Pacati}
\affiliation{Dipartimento di Fisica Nucleare e Teorica, 
Universit\`{a} degli Studi di Pavia and \\
INFN,
Sezione di Pavia, via A. Bassi 6, I-27100 Pavia, Italy}

\date{\today}

\begin{abstract}
The analysis of the recent neutral-current neutrino-nucleus
scattering cross sections measured by the  MiniBooNE Collaboration requires relativistic 
theoretical descriptions also accounting for the role of final state 
interactions. In this work we evaluate differential 
cross sections with  the relativistic distorted-wave 
impulse-approximation and with
the relativistic Green's function model to investigate
the sensitivity to final state interactions.
The role of the strange-quark content of the nucleon form factors is also 
discussed. \end{abstract}


\maketitle


\section{Introduction}
\label{intro}

The MiniBooNE Collaboration has recently reported~\cite{miniboonenc} a 
measurement of the flux-averaged differential cross section as a function of 
the four-momentum transferred squared, $Q^2 = -q^{\mu}q_{\mu}$, for 
neutral-current elastic (NCE) neutrino scattering on CH$_2$ in a $Q^2$ range up to 
$\approx 1.65\ ($GeV/$c)^2$. The neutrino-nucleus NCE reaction in MiniBooNE 
may be considered as scattering of an incident 
neutrino with a single nucleon bound in carbon or free in hydrogen, but 
it can also be sensitive to contributions from collective nuclear 
effects, whose clear understanding is crucial for the analysis of ongoing 
and future  neutrino oscillation 
measurements~\cite{miniboonenc,miniboone,Nakajima:2010fp,Wu08,Lyubushkin:2008pe,minos10}. 
In addition, NCE processes 
can be used to look for strange-quark contributions in the nucleon that may show 
up through the isoscalar weak current. 

Recent results on parity-violating  electron scattering at 
$Q^2 = 0.1\ ($GeV/$c)^2$ ~\cite{Acha:2006my} indicate that the strangeness 
contribution to the electric charge and magnetic moment of the nucleon is 
consistent with zero at 95\% confidence level. In the axial form factor, under 
the usual assumption of a dipole $Q^2-$dependence, the only free parameters 
within the relativistic Fermi gas (RFG) model~\cite{Casper:2002sd,Hayato:2002sd} 
are the nucleon axial mass $M_A$ and the strange-quark contribution 
$\Delta s$, determining the value of the axial form factor at $Q^2 = 0$, 
that is related to the fraction of the nucleon spin carried by the strange 
quark. Recent charged-current quasielastic (CCQE) neutrino-nucleus  
measurements~\cite{miniboone,Gran:2006jn} reported values of 
$M_A \approx \ 1.2-1.3$ GeV/$c^2$, significantly larger than the 
world average value from the deuterium data of 
$M_A = 1.03$ GeV/$c^2$~\cite{Bernard:2001rs,bodek08}. In agreement with these 
new results, the MiniBooNE NCE data provide a best fit for  
$M_A = 1.39 \pm 0.11$ GeV/$c^2$.

A measurement of $\nu (\bar\nu$)-proton elastic scattering at the Brookhaven
National Laboratory (BNL) \cite{bnl} at low $Q^2$ suggested a non-zero value 
for the strange 
axial-vector form factor. However, the BNL data cannot provide us decisive 
conclusions when also strange vector form factors are taken into 
account~\cite{gar}. 
A determination of the strange form factors through a simultaneous analysis of 
$\nu p$, $\bar \nu p$, and $\vec e p$ elastic scattering is performed 
in Ref.~\cite{Pate:2003rk}.

Since cross section measurements are a very
hard experimental task, ratios of cross sections have been
proposed as alternative ways to search for 
strangeness~\cite{gar93,alb02}. 
Moreover, they are expected to be less sensitive to distortion
effects~\cite{Meucci:2004ip,Meucci:2006ir}.
Taking advantage of the fact that at $Q^2 \ge 0.7 \ ($GeV/$c)^2$ single proton events can be
satisfactorily separated from neutron and multiple nucleon events, the MiniBooNE
Collaboration used the ratio of proton-to-nucleon (p/n) cross sections to extract the
strangeness contribution to the axial form factor~\cite{miniboonenc}. 
The resulting value is $\Delta s = 0.08 \pm 0.26$. Although affected by large 
systematic errors because of difficulties in the proton/neutron identification, 
this result is anyhow very interesting, since this is the first
attempt to measure $\Delta s$ using the p/n ratio.
In addition, exploiting its 
data sample of neutrino-nucleus events, the MiniBooNE Collaboration has also 
focused on the neutral-to-charged-current ratio~\cite{miniboonenc,miniboone}, 
that can provide us useful and complementary information. 

The energy region 
considered in the MiniBooNE experiments, with neutrino energy up to 
$\approx \ 3$ GeV and average energy of $\approx \ 0.8$ GeV, requires the use 
of a relativistic model, where 
not only relativistic kinematics should be considered, but also nuclear 
dynamics and current operators should be described within a relativistic 
framework. From the comparison with electron scattering data, the 
RFG turns out to be a too naive model to correctly account for the nuclear 
dynamics, and thus the larger axial mass needed by the RFG could be considered as 
an effective value to incorporate nuclear effects into the calculation. 
Regardless from this question, a comparison between the results of different 
models and the NCE MiniBooNE data is important to clarify the role of the 
various ingredients entering the description of the reaction.

At intermediate energy, quasielastic (QE) electron scattering 
calculations~\cite{Boffi:1993gs,book}, which were able to
successfully describe a wide number of experimental data, can provide a 
useful tool to study neutrino-induced processes. 
However, a careful analysis of $\nu$-nucleus NCE reactions
introduces additional complications, as the final neutrino cannot be measured 
in practice and a final hadron has to be detected: the corresponding cross 
sections are therefore semi-inclusive in the hadronic sector and inclusive in 
the leptonic one. 
Several sophisticated models have been applied in recent years to
$\nu$-nucleus scattering reactions and some of them have been compared with 
the MiniBooNE data, both in the CCQE and in the NCE channels. 
At the level of the impulse approximation (IA), 
models based on the use of a realistic spectral 
function~\cite{Benhar:2010nx,Benhar:2011wy}, 
which are built within a nonrelativistic framework, underestimate the
experimental CCQE and NCE cross sections unless $M_A$ is enlarged with respect
to the world average value. 
The same results are obtained by models based on the relativistic IA 
(RIA)~\cite{Butkevich:2010cr,Butkevich:2011fu,
jusz10}.  
However, the reaction may have significant contributions from effects beyond 
the IA in some kinematic regions where the neutrino flux 
for the experiment has significant strength.  
For instance, in the 
models of 
Refs.~\cite{Martini:2009uj,Martini:2010ex,Martini:2011wp,
Nieves:2011pp,Nieves:2011yp}
the contribution of multinucleon excitations to CCQE scattering 
has been 
found sizable and able to bring the theory in agreement with the experimental
MiniBooNE cross sections without increasing the value of 
$M_A$. 
Fully relativistic microscopic calculations of two-particle-two-hole (2p-2h) 
contributions are very involved and may be bound to model dependent assumptions. 
The part of the 2p-2h 
excitations which may be reached through two-body meson-exchange 
currents (MEC), in particular the contribution of the vector MEC in the 
2p-2h sector, evaluated in the model of Ref.~\cite{DePace:2003xu}, has been 
incorporated in a phenomenological approach based on the superscaling behavior 
of electron scattering data~\cite{Amaro:2010sd,Amaro:2011qb}. 

Within the QE kinematic domain, the treatment of the final-state 
interactions (FSI) between the ejected nucleon and the residual nucleus is an 
essential ingredient for the comparison with data. The relevance of FSI has been
clearly stated in the case of exclusive $(e,e^{\prime}N)$ processes, where the 
use of  complex optical potentials in the distorted wave impulse approximation 
(DWIA) is required~\cite{Boffi:1993gs,book,Udias:1993xy,Udias:1993zs,Meucci:2001qc,
Meucci:2001ty,Radici:2003zz,Tamae:2009zz,Giusti:2011it}. 
In the exclusive reaction, where the final state is completely determined, the
absorptive imaginary part of the optical potential accounts for the flux lost to
different final nuclear states.  In contrast, the analysis of 
inclusive reactions needs all final-state channels to be retained, i.e., the 
flux must be conserved and the DWIA based on the use of an 
absorptive complex potential should be dismissed. 
Different approaches have been used to describe FSI in relativistic calculations for 
the inclusive QE electron- and neutrino-nucleus 
scattering~\cite{Maieron:2003df,Caballero:2006wi,Caballero:2009sn,Meucci:2003cv,
Meucci:2003uy,Meucci:2009nm,Meucci:2011pi}. 
Besides the relativistic plane-wave impulse approximation (RPWIA), where FSI 
are simply neglected, FSI have been included in DWIA calculations where the 
final nucleon state is evaluated with purely real potentials, either retaining 
only the real part of the relativistic energy-dependent optical potential 
(rROP), or using the same relativistic mean field potential considered in 
describing the initial nucleon state (RMF). 
Although conserving the flux, the rROP is unsatisfactory from a theoretical 
point of view, since it relies on  an energy-dependent potential, which 
reflects the different contribution of open inelastic channels for each energy, 
and under such conditions dispersion relations dictate that the potential 
should have a nonzero imaginary term~\cite{hori}.
On the other hand, in the RMF model the same strong energy-independent real 
potential is used for both bound and scattering states. It fulfills the 
dispersion relations~\cite{hori} and also the continuity equation. 

In a different description of FSI relativistic Green's function (RGF) 
techniques~\cite{Capuzzi:1991qd,Meucci:2003uy,Meucci:2003cv,Capuzzi:2004au,
Meucci:2009nm,Meucci:2011pi}
are used. This formalism allows us to reconstruct the flux lost into nonelastic 
channels in the case of the inclusive response starting from the complex optical 
potential which describes elastic nucleon-nucleus scattering data. 
Thus, it provides a consistent treatment of FSI 
in the exclusive and in the inclusive scattering and gives a 
good description of $(e,e')$ data~\cite{Meucci:2003uy,Meucci:2009nm}. 
Moreover, due to the analiticity properties of the optical potential, 
the RGF model fulfills the Coulomb sum rule~\cite{hori,Capuzzi:1991qd,Meucci:2003uy}.

These different descriptions of FSI have been compared in~\cite{Meucci:2009nm} 
for the inclusive QE electron scattering, in~\cite{Meucci:2011pi} for the 
CCQE neutrino scattering, and in~\cite{Meucci:2011vd} with the CCQE MiniBooNE 
data, which are reasonably described by the RGF results without the need to 
increase the value of $M_A$ and are generally underestimated by the other models.

In this paper different relativistic descriptions of FSI for NC $\nu$-nucleus 
reactions in the quasielastic region are presented and compared with the NCE 
MiniBooNE data. 
In these reactions a final nucleon is detected, like in the exclusive 
scattering, but since the final lepton cannot be detected, the final nuclear
state is not completely determined and the cross section includes many possible
final-state channels.
In Refs.~\cite{Meucci:2004ip,Meucci:2006ir,Meucci:2008zz,
Meucci:2006cx,Giusti:2009sy} such a semi-inclusive scattering was
treated with the same relativistic DWIA (RDWIA) model that was successfully
applied to the exclusive $(e,e^{\prime}N)$ reaction, as a process where the 
cross section is obtained from the sum of all the integrated exclusive one-nucleon 
knockout channels. Results for both the semi-inclusive CC and NC cross 
sections were presented and the effects of possible strange-quark contributions 
on the cross sections and on other observables were discussed. In 
RDWIA calculations the imaginary part of the optical potential produces a loss of flux
that accounts for the flux lost in each considered channel towards other 
final channels. Some of these reaction channels are not included in the
experimental cross section when an emitted nucleon is detected and for these
channels this treatment of FSI is correct. There are, however, other channels,
which are excluded by the RDWIA approach but which can contribute to the 
semi-inclusive reaction. Some of these contributions may be small or negligible
for the specific final state considered in the exclusive reaction, but may be 
much more important for all the final states of the semi-inclusive reaction. 
The flux lost towards these channels can be recovered in the RGF, where the 
imaginary part of the optical potental redistributes the strength in all the 
channels and the total flux is conserved.  The RGF, however,
describes the inclusive process and, as such, can include also contributions
of channels that are not included in the cross sections of the semi-inclusive 
reactions. 
Thus, according to the approach adopted to describe FSI, the RDWIA can produce 
cross sections smaller and the RGF larger than the experimental ones. The relevance of the contributions neglected in the RDWIA 
and added in the RGF depends on kinematics. 
It is not easy to disentangle the role of each specific contribution, in particular 
if we consider that both RDWIA and RGF calculations make use of 
phenomenological optical potentials, obtained through a fit of elastic 
proton-nucleus scattering data.

In spite of these uncertainties, the comparison between the results of the RDWIA 
and RGF models with the MiniBooNE data can be helpful for a deeper 
understanding of the role of FSI in the analysis of NCE data.



\section{Results and discussion}
\label{results}

In all the calculations presented in this work the bound nucleon states 
are taken as self-consistent Dirac-Hartree solutions derived 
within a relativistic mean field approach using a Lagrangian containing 
$\sigma$, $\omega$, and $\rho$ mesons~\cite{Serot:1984ey,Sharma:1993it,
Lalazissis:1996rd}. 
The Energy-Dependent but A-Independent EDAI parameterization for the 
complex phenomenological potential of Refs.~\cite{chc,Cooper:1993nx}, which is 
fitted to elastic proton-$^{12}$C scattering data, has been used. 
The Energy-Dependent and A-Dependent EDAD1 parameterization, which is fitted to 
elastic proton scattering data on several nuclei in an energy range 
up to 1040 MeV, has also been used for some calculations.
We note that whereas EDAD1 is a global parameterization, EDAI is a 
single-nucleus parameterization, which is 
constructed to better reproduce the elastic proton-$^{12}$C 
phenomenology~\cite{Cooper:1993nx}, and also leads to CCQE results in 
better agreement with the MiniBooNE data~\cite{Meucci:2011vd}. 

\begin{figure}[t]
\begin{center}
\includegraphics[scale=0.45]{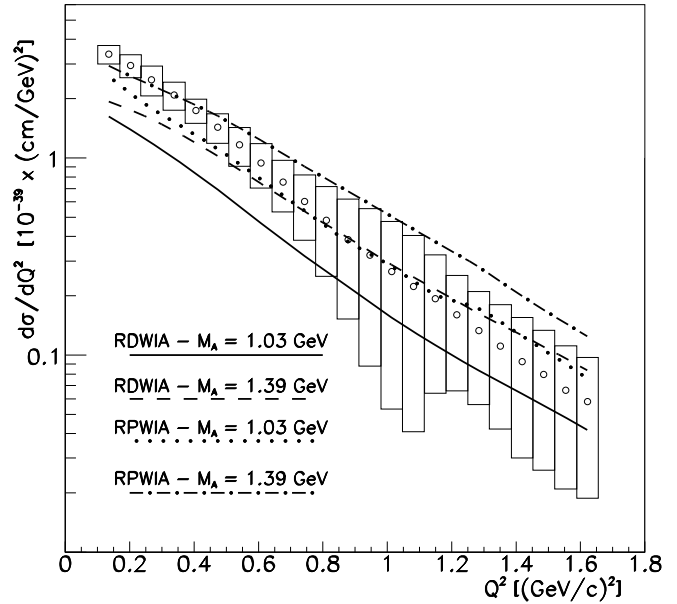} 
\end{center}
	\caption{NCE flux-averaged 
	$\left ( \nu N \longrightarrow \nu N \right )$ cross section 
	as a function of $Q^2$ calculated with two different values of the
	axial mass in the RDWIA (solid and dashed lines) and RPWIA 
	(dotted and dot-dashed lines).
	The data are from MiniBooNE~\cite{miniboonenc}. 
	\label{f1}}
	\end{figure}
In Fig.~\ref{f1} the NCE $(\nu N \longrightarrow \nu N)$ cross sections 
averaged over the neutrino flux are shown as a function of $Q^2$ and compared 
with the MiniBooNE data~\cite{miniboonenc}. Calculations are performed in the
RPWIA and RDWIA with 
$M_A = 1.03$ and $1.39$ GeV/$c^2$ and neglect possible strange-quark effects.
A larger value of $M_A$ gives a larger cross section, because of the dominant 
role played by the axial-vector current in NC scattering, but the enhancement is 
not linearly proportional to $M_A$ and, therefore, also the shape of the 
cross section is slightly modified. 
In the comparison with data, the RPWIA results show a generally 
satisfactory, although not perfect, agreement with the magnitude, while some 
differences are obtained with respect the shape of the experimental cross
section. In the RPWIA, however, FSI are completely neglected.  
The RDWIA results are smaller than the RPWIA ones and therefore also smaller 
than the experimental data. This is due to the imaginary part of the optical 
potential, that in the RDWIA gives an absorption that reduces the calculated cross section.
In Fig.~\ref{f1} the RDWIA calculations with the EDAI potential   
generally underestimate the NCE cross section, unless $Q^2 \ge \ 0.8\ ($GeV/$c)^2$
for $M_A = 1.03$ GeV/$c^2$ and $Q^2 \ge \ 0.6\ ($GeV/$c)^2$ for $M_A = 1.39$ 
GeV/$c^2$. 
We have checked that close RDWIA results are obtained with the EDAD1 potential, but 
for small values of $Q^2$, where there are visible differences that can be 
attributed to the different imaginary parts of the two optical potentials 
in the low energy region.

\begin{figure}[t]
\begin{center}
\includegraphics[scale=0.45]{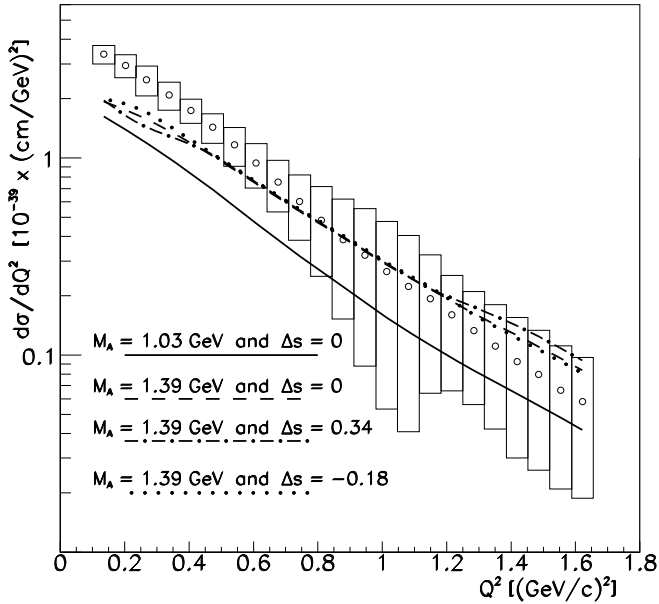} 
\end{center}
	\caption{NCE flux-averaged 
	$\left ( \nu N \longrightarrow \nu N \right )$ cross section 
	as a function of $Q^2$, calculated in the RDWIA with three different 
	values of $\Delta s$ and $M_A = 1.39$ GeV$/c^2$. Results with $\Delta s = 0$
	and $M_A = 1.03$ GeV$/c^2$ are also shown. The data are from
	MiniBooNE~\cite{miniboonenc}. 
	\label{f2}}
	\end{figure}
In Fig.~\ref{f2} we show our RDWIA results with $\Delta s = -0.18$ and 
+0.34 and $M_A = 1.39$ GeV/$c^2$. These are the upper and lower limits for 
$\Delta s$ found by MinibooNE~\cite{miniboonenc}. 
The results for $\Delta s = 0$ with $M_A = 1.03$ and
$1.39$ GeV/$c^2$ are also shown again for a comparison. 
The MiniBooNE NCE cross section is nearly
independent of $\Delta s$, as the combined effects on proton 
and neutron events almost cancel. In the calculations a negative $\Delta s $ 
produces an enhancement of the proton and a suppression of the neutron 
contribution, which are approximately of the same size (see 
also~\cite{Benhar:2011wy}). In the case of positive $\Delta s $ the 
effect is reversed. As a consequence, the effects of different values of  
$\Delta s = 0$ are minimal and smaller than the uncertainties due to $M_A$, 
whose value has a visible impact on the MiniBooNE NCE cross section.

\begin{figure}[t]
\begin{center}
\includegraphics[scale=0.45]{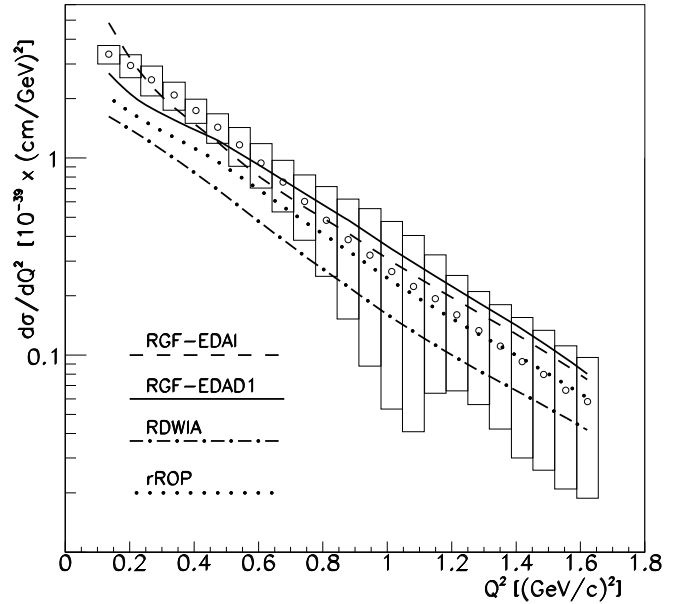} 
\end{center}
	\caption{NCE flux-averaged 
	$\left ( \nu N \longrightarrow \nu N \right )$ cross section 
	as a function of $Q^2$ calculated 
	with the RGF-EDAD1 (solid line) and RGF-EDAI (dashed line). The 
	dotted and dot-dashed lines are rROP and RDWIA results calculated with 
	the EDAI potential, respectively.
	The data are from
	MiniBooNE~\cite{miniboonenc}. 
	\label{f3}}
	\end{figure}
In Fig.~\ref{f3} we show our RGF results calculated with both  EDAI and
EDAD1 potentials and compare them with the RDWIA and the rROP results calculated
with the EDAI potential. All these calculations have been performed with 
$M_A = 1.03$ GeV/$c^2$ and $\Delta s = 0$. 
The RGF cross sections with both optical potentials are larger than the RDWIA
and the rROP ones. The rROP result, where a purely real optical potential is 
used, underestimates the experimental cross section  for 
$Q^2 \le \ 0.6\ ($GeV/$c)^2$. We observe that a rROP calculation with a larger
axial mass, e.g., $M_A \approx 1.3 - 1.4$ GeV/$c^2$, is able to reproduce the
data with good accuracy. However, we note that, independently of its result in
comparison with data, the rROP model, which is based on 
an energy-dependent potential, has important physical drawbacks.
The RDWIA cross section with the EDAI 
potential is the same already presented in Fig.~\ref{f1} and gives 
in Fig.~\ref{f3} the lowest cross section. The differences between
the RGF results calculated with the two optical potentials are clearly visible, 
although not too large, the RGF-EDAI cross section being in good agreement 
with the shape and the magnitude of the experimental cross section, and the 
RGF-EDAD1  below the data only at the smallest values of $Q^2$ considered in the
figure. The differences between the RGF results obtained with the two 
phenomenological optical potentials can give an idea of how uncertainties in 
the determination of this important ingredient can affect the predictions of 
the model. These differences are basically due to the
different values of the imaginary parts of the two potentials, particularly for 
the energies considered in kinematics with the lowest values of $Q^2$. 
The real terms of the optical potentials are very similar for different 
parameterizations and 
give very similar results. In the rROP calculation shown in the figure the real
part of the EDAI potential has been used, but a calculation with EDAD1 would
give in practice the same result.

The results displayed in Fig.~\ref{f3} emphasize the important role of FSI and
in particular of the imaginary part of the relativistic optical potential, that
plays a different role in the different approaches. In the rROP the 
imaginary part is neglected. In the RDWIA it gives an absorption that accounts
for the flux lost in each channel towards other channels that are not included in
the model. In the RGF the imaginary part redistributes the flux in all the
final-state channels: in each channel it accounts for the flux lost towards
other inelastic channels and recovered  for the inclusive scattering making use
of the dispersion relations. The results obtained in the different 
models give an idea of the relevance of these contributions. The larger cross
sections in the RGF arise from the translation to the inclusive strength of the
overall effect of inelastic channels. We have already noticed, however, that 
while the RDWIA neglects the contributions of some channels that can be 
included in the semi-inclusive reaction where only the emitted nucleon is 
detected, the RGF is appropriate for the inclusive scattering where only the 
final lepton is detected, and can take into account also contributions that are not 
included in the semi-inclusive process.
From this point of view the RDWIA can represent a lower limit and the RGF 
an upper limit to the calculated NCE cross sections. 

\begin{figure}[t]
\begin{center}
\includegraphics[scale=0.45]{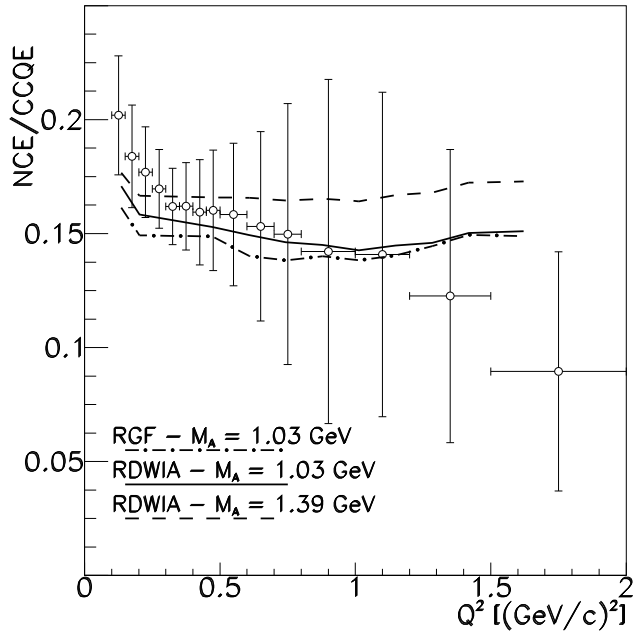} 
\end{center}
	\caption{The ratio  of NCE/CCQE cross sections 
	as a function of $Q^2$ calculated in the RDWIA with $M_A = 1.03$ (solid
	line) and
	$1.39$ GeV$/c^2$ (dashed line), and in the RGF with  $1.03$ GeV$/c^2$ 
       (dot-dashed-line).
	The data are from
	MiniBooNE~\cite{miniboonenc}.
	\label{f4}}
	\end{figure}
The MiniBooNE Collaboration reported results for the flux-averaged differential
cross section both in the CCQE and in the NCE scattering. In Ref.~\cite{miniboonenc} these
results are used to extract the NCE/CCQE ratio as a function of $Q^2$. 
This ratio can be useful to compare the results of the two measurements. 
In Fig.~\ref{f4} we show our results for the NCE/CCQE
ratio. We note that both NCE and CCQE cross sections are per target nucleon,
thus there are 14/6 times more target nucleons in the numerator than in the 
denominator~\cite{miniboonenc}.
The results of the RGF and RDWIA with the EDAI potential are similar 
and within the experimental errors. This is in accordance with the
observation that ratios of cross sections do not depend on FSI effects.
The RDWIA model, which gives much lower predictions for the cross sections than
the RGF, can produce results for the ratio close to the RGF ones and in  
overall agreement with the data.
A more significant effect is given by a larger $M_A$. 
When $M_A = 1.39$ GeV/$c^2$ the NCE/CCQE ratio is enhanced up to $\approx 30\%$ . 
This means that the axial mass has a different role in the NC and in the CC
scattering.

\begin{figure}[t]
\begin{center}
\includegraphics[scale=0.45]{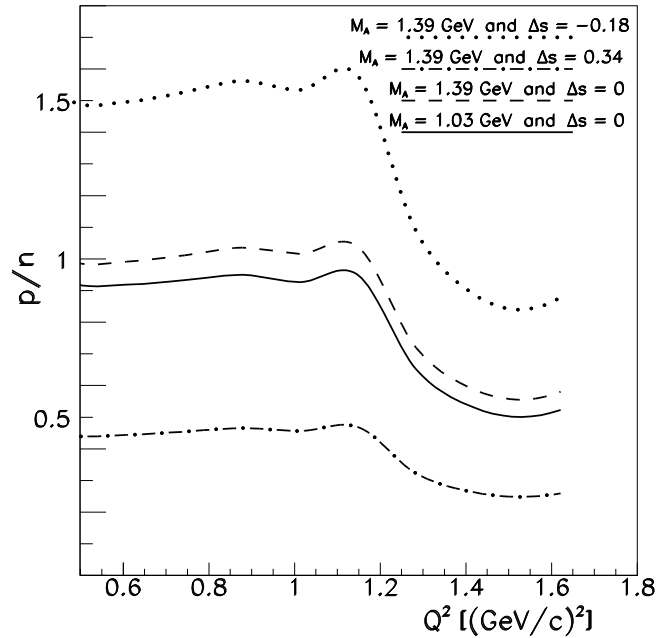} 
\end{center}
	\caption{The ratio  of p/n cross sections 
	as a function of $Q^2$ calculated in the RDWIA with three different values 
	of $\Delta s$ and $M_A = 1.39$ GeV$/c^2$. Results with $\Delta s = 0$
	and $M_A = 1.03$ GeV$/c^2$ are also shown. 
	\label{f5}}
	\end{figure}
In order to measure the
strangeness contribution to the axial form factor the MiniBooNE
Collaboration used the ratio of p/n cross sections for protons 
above the Cherenkov threshold as a function of the reconstructed nucleon 
kinetic energy~\cite{miniboonenc}. 
In Fig.~\ref{f5} we show our RDWIA results for the p/n ratio. 
This ratio was proposed as an efficient way to measure 
$\Delta s $~\cite{gar93,alb02} as the distortion effects should be largely 
reduced, but was later given up
due to the difficulties associated with neutron detection.
The p/n ratio is very sensitive to the 
strange-quark contribution, as the axial-vector strangeness $\Delta s $ 
interferes with the isovector contribution to the axial form factor $g_A = 1.26$
 with one sign in the 
numerator and with the opposite sign in the denominator. 
We already investigated in Ref. \cite{Meucci:2006ir} the sensitivity of 
the p/n ratio to  $\Delta s $ as well as to the strange-quark contribution to 
the vector form factors. A large dependence on $\Delta s $ was obtained, but also
the effect of the magnetic strangeness was significant.
However, we note that the p/n ratio of Fig.~\ref{f5} is obtained by dividing the
flux-integrated cross sections with one proton or one nucleon in the
final state and, moreover, that in the CH$_2$ target there are 8 protons and 
6 neutrons. Thus, it is not straightforward to compare the results in 
Fig.~\ref{f5} with those of Ref.~\cite{Meucci:2006ir}. 
 Because of the independence of the p/n ratio on FSI, the 
results with the RGF and rROP models are similar to the RDWIA ones and are not 
shown in the figure. 
The ratio is largely 
enhanced when a negative $\Delta s $
is included and suppressed when a positive $\Delta s $ is considered. 
Varying the axial mass modifies the ratio up to $\approx 10\%$.  


\section{Summary and conclusions}
\label{conc}

In this paper we have compared the predictions of 
different relativistic descriptions of FSI for quasielastic 
NC neutrino-nucleus scattering with the MiniBooNE NCE data. 
In the RPWIA FSI are simply neglected, in the rROP they are described retaining only
the real part of the relativistic energy-dependent optical potential, while in
the RDWIA and in the RGF the full complex optical potential, with its real and imaginary
parts, is used to account for FSI.
  
The RDWIA is based on the same model that was widely and successfully applied 
to the analysis of the exclusive $(e,e^{\prime}p)$ knockout reaction, where 
the final state is completely determined. In the exclusive reaction the 
absorptive imaginary part of the optical potential, which accounts for the 
flux lost in the considered elastic channel to all inelastic final-state 
channels, gives a reduction of the calculated cross section that is required 
for a proper description of the experimental data. In the RDWIA the 
cross section for the semi-inclusive reaction where only the emitted nucleon is 
detected is obtained from the sum of all the integrated exclusive one-nucleon 
knockout channels. In this case, however, the
reduction produced in each channel by the imaginary part of the optical
potential, that can be appropriate for the exclusive reaction, neglects some 
final-state channels that can contribute to the semi-inclusive reaction.   
All final-state channels are included in the RGF, where the flux lost in 
each channel is recovered in the other channels just by the imaginary part of
the optical potential and the total flux is conserved. The RGF model is 
appropriate for the inclusive scattering, where only the outgoing lepton is 
detected, and with the use of the same complex optical potential it provides a
consistent treatment of FSI in the exclusive and in the inclusive process. In
comparison with data, the RGF is able to give a good description of the
$(e,e')$ experimental cross sections in the QE region and also of the recent 
CCQE MiniBooNE data without the need to increase the standard value of the
axial mass.
The application of the RGF to the semi-inclusive NCE scattering can 
recover important contributions that are omitted in the RDWIA, and can give, from the
comparison with the DWIA results, an indication of their relevance. It can 
also include, however, contributions of channels which are present only in the inclusive 
but not in the semi-inclusive reaction.  

The RPWIA, rROP, and RDWIA results generally underpredict the MiniBooNE NCE 
data, the RDWIA giving the lowest cross section, unless the standard value of 
$M_A$ is significantly enlarged. In contrast, the RGF results are in reasonable
agreement with the experimental NCE cross section without the need to increase
the standard value of $M_A$. 

The enhancement of the RGF cross section can be ascribed to the contribution of reaction
channels that are not included in the other models. It can be due, for
instance, to re-scattering processes of the nucleon in its way out of the
nucleus, to non-nucleonic $\Delta$ excitations, which may arise during nucleon
propagation, with or without real pion production, as well as to multinucleon 
processes. These contributions are not included explicitly in the model with a
microscopic approach, but can be recovered, to some extent, in the RGF by the 
imaginary part of the phenomenological optical potential. With the use of such a
phenomenological ingredient, however, we cannot disentangle the role of 
different reaction processes and explain in detail the origin of the recovered 
strength.
 
If all these contributions can be present in the inclusive scattering, the role
of multinucleon processes in the NCE experimental data is not clear. 
It is a fact, however, that the theoretical analysis of MiniBooNE CCQE and NCE 
data presents a common problem: not only the simple RFG, but also other models, based on the IA and
including only one-nucleon knockout contributions, require a larger value of 
$M_A$ to reproduce the magnitude of the experimental 
cross sections.
The calculations required for the theoretical analysis must consider the entire 
kinematical range of the relevant MiniBooNE neutrino energies. Additional 
complications may arise from the flux-average procedure to evaluate the CCQE and
NCE cross sections, which implies a convolution of the
double differential cross section over the neutrino spectrum. 
It has been argued~\cite{Benhar:2010nx,Benhar:2011wy} that, due to the 
uncertainties associated with the flux-average procedure, the MiniBooNE cross
sections can include contributions from different kinematic regions, where 
other reaction mechanisms than one-nucleon knockout are known to be dominant. 
Moreover, further ambiguities arise for the MiniBooNE NCE cross section,
which is given in bins where $Q^2$ is reconstructed from the kinetic 
energies of the ejected nucleons. 

Models including other contributions that one-nucleon knockout, like our RGF,
but also the model of 
Refs.~\cite{Martini:2009uj,Martini:2010ex,Martini:2011wp}, where multinucleon
components are explicitly included, are able to describe both the MiniBooNE 
CCQE and NCE data without the need to change the value of
the axial mass. The two models are different, but they seem to go in the same
direction. In the RGF, however, the enhancement of the cross section cannot be
attributed only to multinucleon processes, since we cannot disentangle 
the role  of the various contributions included in the phenomenological 
optical potential. 
  
In order to clarify the content of the enhancement of the RGF cross sections 
compared to those of the IA models, a careful evaluation of all nuclear effects 
and of the relevance of multinucleon emission and of some non-nucleonic 
contributions~\cite {Leitner:2010kp} is required.   
The comparison with the results of the RMF model, where only the purely 
nucleonic contribution is included, would be interesting for a deeper
understanding.
Processes involving two-body currents, whose importance has been discussed in
Refs. \cite{Benhar:2011wy,Amaro:2010sd,Amaro:2011qb,bodek11}, should also be
taken into account explicitly and consistently in a model to clarify the role of multinucleon
emission. 

The RGF predictions are also affected by uncertainties in the determination of
the phenomenological optical potential. 
At present, lacking a phenomenological optical potential which exactly 
fullfills the dispersion relations in the whole energy region of interest, 
the RGF prediction is not univocally determined from the elastic phenomenology.
The differences between the RGF results obtained in the present investigation 
with the EDAI and  EDA1 potentials are visible, although smaller than for the 
CCQE cross sections in Ref.~\cite{Meucci:2011vd}. 
These differences are produced by the different imaginary part, that is the 
crucial ingredient in both RDWIA and RGF calculations, the real part
being very similar for different parametrizations of the optical potential. 
It is interesting to notice that the best predictions in comparison with both 
CCQE and NCE data are given by the EDAI potential, that is also able to give 
the better description of the elastic proton-$^{12}$C phenomenology. 
A better determination of a phenomenological relativistic optical potential, 
which closely fullfills the dispersion relations, would be anyhow desirable and 
deserves further investigation.

In this work we have investigated also the role of a possible strange-quark 
contribution $\Delta s$ to the axial nucleon form factor. The calculated cross
sections are almost unaffected by $\Delta s$, as the combined effects of $\Delta s$
on proton and neutron events almost cancel. As a consequence, also the ratio of
NCE/CCQE cross sections is unaffected by $\Delta s$ if both proton and
neutron events are included in the NCE cross section. Experimental information
on the strange-quark contribution to the NCE/CCQE ratio can be obtained if only proton (or
neutron) emission is considered. The p/n ratio is very sensitive to the
strange-quark contribution, but requires the explicit separation of the proton
and neutron cross sections.


\begin{acknowledgments}

We are grateful to Franco Capuzzi for useful discussions.  This work was 
partially supported by the Italian MIUR through the PRIN 2009 research project.

\end{acknowledgments}


\bibliography{rif-nc}

\end{document}